\documentclass{ws-jai}
\usepackage[flushleft]{threeparttable}
\begin{document}

\catchline{}{}{}{}{} 

\markboth{Dominguez-Tagle}{First light of the Integral Field Unit of GRIS}

\title{First light of the Integral Field Unit of GRIS on the GREGOR solar telescope}

\author{C. Dominguez-Tagle$^{1,2 *}$, M. Collados$^{1,2}$,   R. Lopez$^{1,2}$, J. J. Vaz Cedillo$^{1,2}$, M.A. Esteves $^{3}$, O. Grassin $^{3}$, \\  N. Vega$^{1,2}$,  A. Mato$^{1,2}$, J. Quintero$^{1,2}$,  H. Rodriguez $^{1,2}$, S. Regalado$^{1,2}$,   F. Gonzalez$^{1,2}$}

\address{
$^{1}$Instituto de Astrofisica de Canarias, C/Via Lactea, 38205 Tenerife, Spain.\\
$^{2}$Departamento de Astrofisica, Universidad de La Laguna, 38206, Tenerife, Spain\\
$^{3}$Leibniz-Institute f\"{u}r Sonnenphysik, Sch\"oneckstrasse 6, 79104 Freiburg, Germany
}

\maketitle

\corres{$^{*}$Corresponding author.}

\begin{history}
\received{(to be inserted by publisher)};
\revised{(to be inserted by publisher)};
\accepted{(to be inserted by publisher)};
\end{history}

\begin{abstract}

An Integral Field Unit (IFU) based on image-slicers has been added to the GREGOR Infrared Spectrograph (GRIS). 
This upgrade to the instrument makes possible 2D spectropolarimetry in the near-infrared by  simultaneously recording the full Stokes profiles of spectral lines (in a given spectral interval) at all the points in the field of view. It provides high-cadence spectropolarimetric observations at the instrument's high spatial resolution and high polarization sensitivity at the GREGOR solar telescope.  
The IFU is ideal for observing the polarized spectrum of fast-evolving solar features at high spatial and spectral resolutions. The high observing cadence opens the possibility of time-series observations. The analysis of observations to this level of accuracy is essential for understanding the complex dynamics and interactions of solar plasma and magnetic fields. 
 The image slicer of the IFU has  eight slices of width 100~$\mu$m, covering a total field of view of 6$^{\prime\prime}$~$\times$~3$^{\prime\prime}$. 
 It was designed and built within the framework of the European projects SOLARNET and GREST,  as a prototype for future instruments of the European Solar Telescope (EST) and  was integrated into GRIS. After two commissioning campaigns in 2017 and 2018, the IFU was finally installed at the end of September 2018  and offered to all observers who use the telescope.
\end{abstract}

\keywords{Integral Field Unit; IFU; GREGOR; GRIS; spectropolarimetry; solar.}

\section{Introduction}

\noindent Integral field units (IFUs) based on image-slicers are commonly used in astrophysical night-time slit-spectroscopy. They were initially proposed by \citet{Bowen38} and have been developed since the work of \citet{Weitzel96} and \citet{Content97}. The  working principle of image slicers is to slice a region of the focal-plane image into a series of small rectangles and  rearrange them in the form of a long slit. This slit is composed of a given number of mini-slits (equal to the number of slicer mirrors). The width of the individual slicer mirrors determines the slit width that feeds the spectrograph, into which the IFU is integrated. 

 In solar physics, IFUs based on this technology can be used to record fast-evolving solar features at high spatial and spectral resolution, even at near-infrared (NIR) wavelengths. 
  As described in a review of instrumentation for solar spectropolarimetry by \citet{Iglesias19}, there are few instruments  currently in operation \citep[see][]{SPINOR,FIRS,NIRIS,Collados12,jaeggli22} that can perform solar spectropolarimetry in the NIR ({\itshape i.e.}, up to 1.7 $\mu$m). 
  
 In the case of GRIS \citep[GREGOR Infrared Spectrograph, see][]{Collados12}, adding an IFU to the instrument makes possible to obtain spectra at all the points in a 2D field of view (FOV) and get the full Stokes profiles of
 the spectral lines in a given spectral interval. This makes possible to observe, at high spatial and spectral resolutions, the polarized spectrum of fast-evolving solar features at a cadence of a few seconds. 
 The  IFU added to GRIS was designed at the Instituto de Astrof\'{i}sica de Canarias (IAC) as a 
 first approach to
 the conceptual instrument called ``MUSICA'' \cite{Calcines13b},  for EST instrumentation \cite{Calcines13}.
 The IFU was built in collaboration with Winlight Systems (formerly Winlight Optics) within the framework of the  SOLARNET and Getting Ready for EST (GREST) European projects. The IFU was taken  to the GREGOR solar telescope \cite{gregor12} and integrated into GRIS in 2017. From the first tests, it was clear that some work  still needed to be done, and a second commissioning campaign was carried out in 2018 \cite{cdt18}. During the scientific validation' procedure, a number of spectropolarimetric measurements at the solar disk center were recorded. 
 After these campaigns, the IFU in GRIS was made available to the entire GREGOR observing  community. 

Between 2018 and 2019, more than 15 scientific teams observed with the IFU. It was  in use for more than 70\% of the telescope's observing time. More than 100 days of scientific data have been archived (i.e., selecting those obtained in good observing conditions). In 2020 and the first semester of 2021, the IFU was not offered because of the pandemic and changes in the telescope optics. Four scientific articles with results based on IFU  observations have already been published \cite{Anjali20,campbell21,tetsu21,nelson21}. 

The instrument and the IFU are described in Section~\ref{cdt:instdes}, including an analysis of performance. Some first light science observations and other examples are presented in Section~\ref{cdt:sciobsv}. Our conclusions are given in Section~\ref{cdt:concl}.

\section{Instrument description}
\label{cdt:instdes}

\subsection{GRIS}

GRIS \cite{Collados12} is  a spectropolarimeter consisting of  a 6~m focal-length  Czerny--Turner spectrograph and a polarimeter built from the heritage of TIP-II \cite{Collados07}.
 GRIS works in the wavelength range 1.0--2.3~$\mu$m. Its 1k $\times$ 1k NIR detector can be read at up to 36~fps. The polarimeter has two sets of Ferroelectric Liquid Crystals, one optimized to work at 1.0--1.3 $\mu$m and the other at 1.5--1.8~$\mu$m. GRIS has two  observing modes,  spectroscopy and spectropolarimetry, both with a long slit (0.26$^{\prime\prime}$~width~$\times$~60$^{\prime\prime}$ length). It has a Slit Scan Unit (SSU) to do scans perpendicular to the slit. The SSU  allows the scanning of a FOV up to 60$^{\prime\prime}$~$\times$~64$^{\prime\prime}$ to be covered. The polarimeter is placed immediately after the SSU. As described in  \citet{Collados12}, the slit mask in the SSU is inclined $15^{\circ}$ with respect to the entrance beam in order to reflect the light that arrives out of the slit towards 
 a slit-jaw imaging system.
  This inclination does not alter the operation of the spectrograph and allows  a field context image to be generated. 
 Table~\ref{cdt:spectable} shows the specifications of the slit mode in the first column. The IFU mode is discussed in the next section.

\begin{wstable}[ht]
      \caption{GRIS slit and IFU specifications}
         \label{cdt:spectable} 
	 \begin{tabular}{l l p{0.5\linewidth}} \toprule
            Slit mode           &   IFU mode \\ \colrule
            Number of slits: 1   & Number of slices: 8 \\
	    Slit length: $<$ 60$^{\prime\prime}$ & Slice length: 6$^{\prime\prime}$ \\
	    Slit width:  0.26$^{\prime\prime}$   & Slice width: 0.375$^{\prime\prime}$ (100 $\mu$m) \\ 
            Slit Scan (SSU)           & 2D Scan (FOVSS) \\
	    Moves in 1 direction & Moves in 2 directions \\
	    Max. FOV: $60^{\prime\prime} \times 64^{\prime\prime}$  & Max. FOV: $60^{\prime\prime}\times 60^{\prime\prime}$ \\ 
	    Double sampling mode:& \\
	    half slit width (see Section~\ref{cdt:sciobsv})      & same  \\ 
	    Spectral wavelength ranges: & \\
	    18 \AA $\:$ at 10830 \AA 	 & same\\
	    40 \AA $\:$ at 15650 \AA 	 & same\\	    
	    Pixel scale: 0.135$^{\prime\prime}$   & same \\
	    Simultaneous slit-jaw images acquisition 	 & same \\
	    Detector: 1k $\times$ 1k	 & same \\
	    Well depth: 16384 ADUs (14 bits)\\
	    Gain: 19 e-/ADU & same\\ \botrule

	 \end{tabular}
\end{wstable}

\subsection{IFU}
\label{cdt:ifu}

\begin{figure}[ht]
     \begin{center}
   \includegraphics[width=15cm]{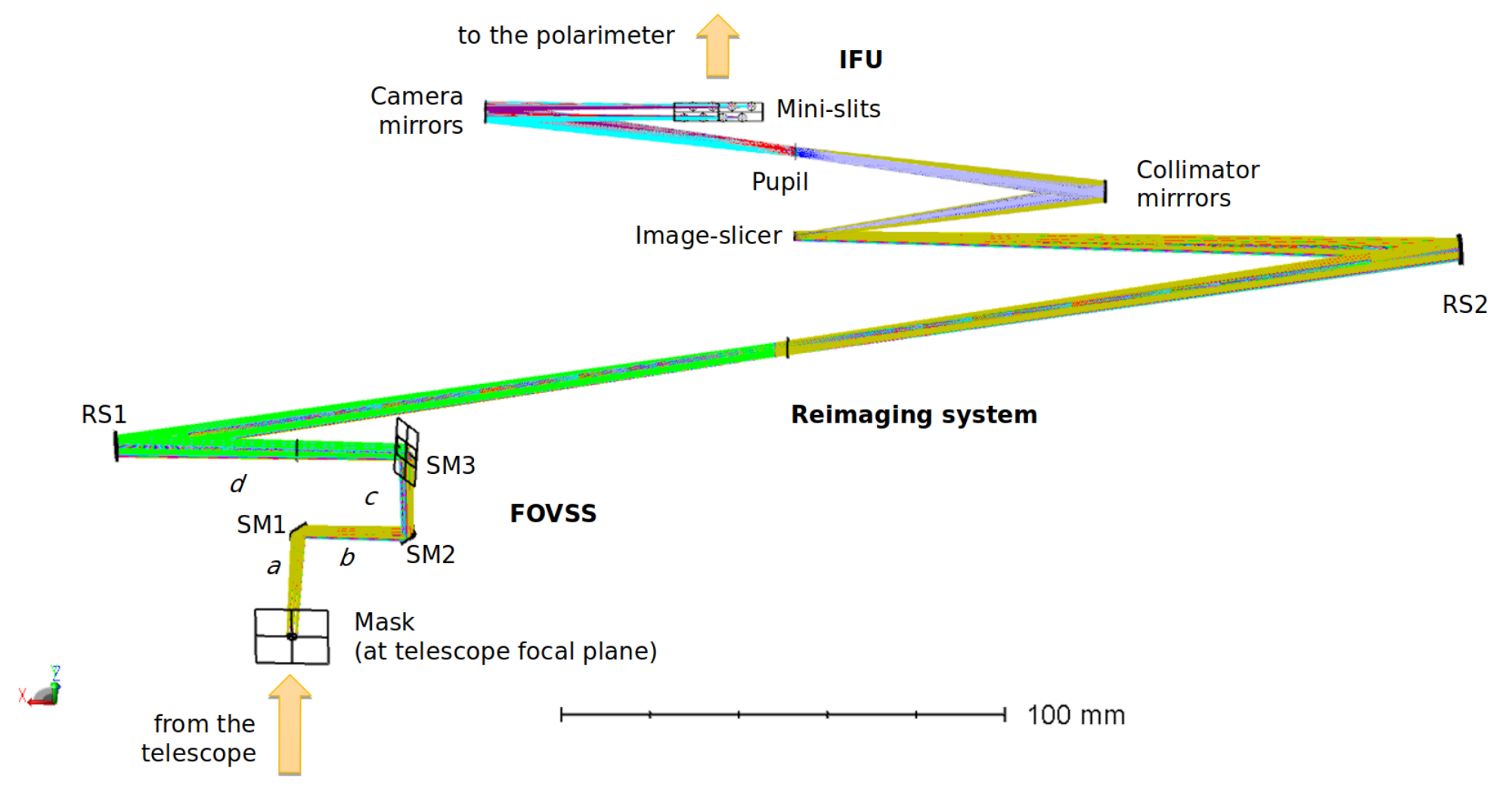}
      \caption{Light path in the IFU mode. The telescope focus is at the bottom of the figure (Mask). The light goes through the FOVSS and the reimaging system up to the IFU and then continues towards the polarimeter. The light path distances ($a, b, c, d$) in the FOVSS are shown next to its components, the entrance mask and the three SMs. The components for the reimaging system are the two RS mirrors. Those for the IFU are the image slicer and the arrays of collimator and camera mirrors. The position of the output mini-slits is also shown. }
         \label{cdt:path}
	 \end{center}
   \end{figure}

The IFU has been coupled to GRIS to add an additional observing mode. Currently, it is possible to exchange the slit and IFU  modes in a single day \cite{Vega16}. In the slit mode, the slit is placed at the focal plane of the telescope and is part of the SSU. The light path exits the SSU and goes through the polarimeter to the rest of the spectrograph. In IFU mode, the arrangement is different because of the available space and the size of the optical components of the polarimeter. Figure~\ref{cdt:path} shows the layout of the IFU mode. An entrance mask is placed at the telescope's focal plane to limit the FOV to the size of the image slicer. The mask is the first element of the Field of View Scan System (FOVSS). This new scanning system is described in Section~\ref{cdt:scans}. A reimaging system is incorporated in order to place the IFU in a new focal plane. The reimaging system is based on a classic collimator-camera design and includes two mirrors (RS1 and RS2). The image slicer of the IFU (shown at the top of Figure~\ref{cdt:path}) is at the focal plane generated by the RS2 mirror.

\begin{figure}[ht]
     \begin{center}
   \includegraphics[width=13cm]{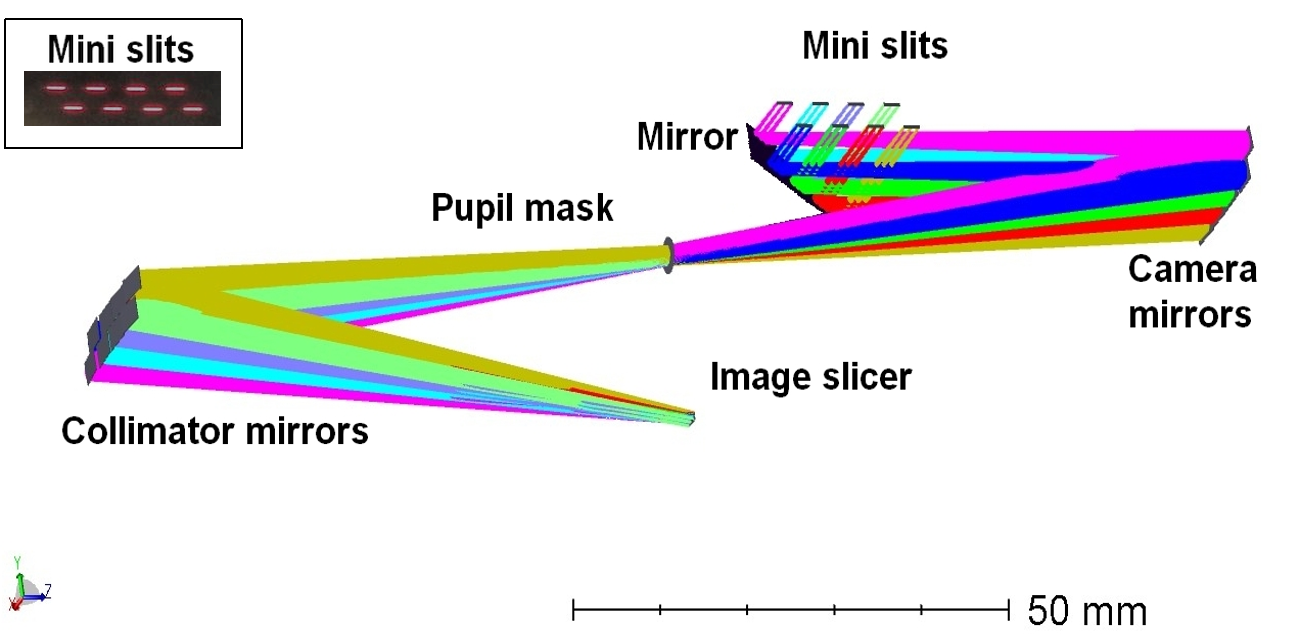}
      \caption{IFU optical layout. The light paths generated by every slice of the image are represented in different colors. The insert shows a real image of the output mini-slits.  The layout orientation is left-right reversed, compared to that in Figure~\ref{cdt:path}, in order to show the mini-slits projected towards the reader. }
         \label{cdt:figzem}
	 \end{center}
   \end{figure}

 The IFU covers a FOV of $6^{\prime\prime}\times 3^{\prime\prime}$ in a single exposure. 
It is based on image-slicer technology with eight slicer mirrors of size 1.8~mm~$\times$~0.1~mm, each. These mirrors cut a rectangular region of the image at the focal plane and reorganize it into a long slit formed by eight mini-slits. The whole IFU body is fabricated in Zerodur to reduce thermal sensitivity because the instrument works at room temperature. The reorganization of the image  from the image slicer  towards the  output slit is accomplished by a reimaging system, which has a classic collimator-camera design. There is a pair of collimator and camera mirrors for each slice. The optical design evolved from the U-path described in \citet{Calcines14} to a Z-path, using spherical mirrors for the collimator and camera. The system is designed  to be telecentric and includes a mask at the pupil plane to reduce straylight. The eight beams have the pupil in the same position so that one pupil mask works for all of them.  The output slit is arranged in two rows of mini-slits in order to minimize geometrical distortion and optical aberrations. The optical design is shown in Figure~\ref{cdt:figzem}. The insert shows the output mini-slits illuminated with sunlight. There is red light at the mini-slits because the IFU is fed by wavelengths over 650~nm.  

\begin{figure}
\begin{center}
    \includegraphics[height=5cm]{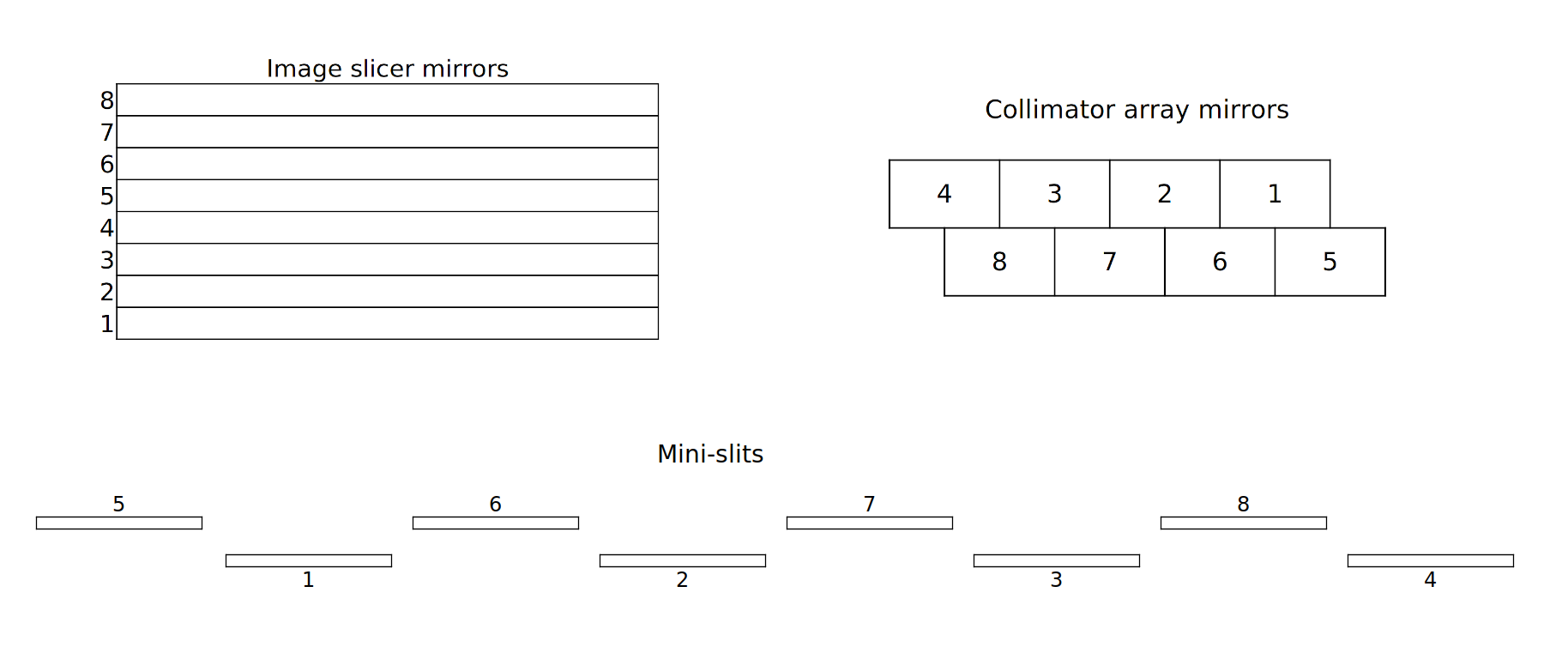}
\end{center}
\caption {Slicer, collimator and mini-slit numbering for identification purposes.}
\label{cdt:numslice}  
\end{figure}

The collimator and camera mirrors are grouped into arrays and their numbering is shown in  Figure~\ref{cdt:numslice}. It can be seen that the central mirrors in the image slicer correspond to the edges of the collimator and mini-slit arrays. After the  mini-slits, the light path continues to the polarimeter and the rest of the spectrograph. Two flat mirrors are inserted in the path after the polarimeter in order to take the beam back to the spectrograph optical axis and to
  compensate for the difference in the path length with respect to the slit mode.

From an operational point of view, observations in IFU mode are conducted  in a similar way to those in slit mode, with the exception that the IFU provides the opportunity to observe a 2D region in one shot.
In addition, it has scanning capabilities with the FOVSS to record a larger solar area. 
Both the  SSU and the FOVSS can cover roughly the same area, although the IFU is most useful for fast small-area scans. The second column of   Table~\ref{cdt:spectable} shows the specifications of the IFU mode.

   \begin{figure}[ht]
     \begin{center}
    \includegraphics[width=8cm]{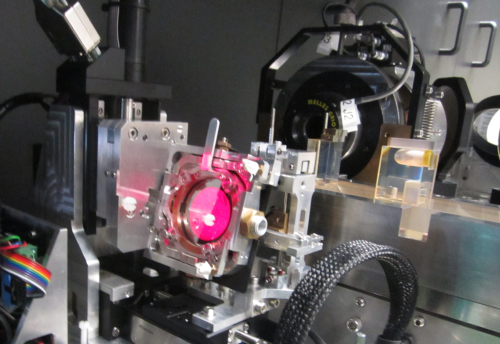}
    \includegraphics[width=8cm]{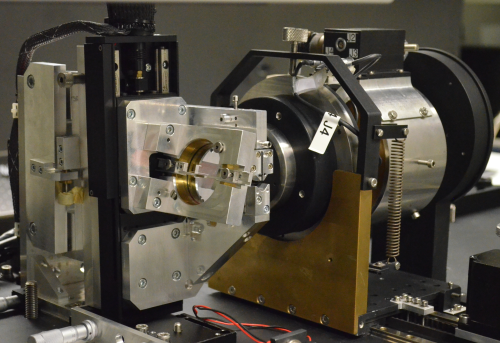}
      \caption{Pictures of the entrance mask in the  IFU mode (left)  and the slit mode (right). The IFU mode is showed with sunlight over the mask. Here the FOVSS, the IFU (Zerodur block) and the polarimeter (metallic cylinder) are partially visible. In the slit-mode picture the SSU and the polarimeter are clearly visible.}
         \label{cdt:pcicssu}
	 \end{center} 
   \end{figure}

As described before, the entrance mask defines the FOV with precision in order to illuminate only the useful part of the image slicer. 
The mask is placed at the entrance of the FOVSS (see Figure~\ref{cdt:pcicssu}) and has the same inclination  as the slit mask ($15^{\circ}$). During the first campaigns it was a rectangle cut on the aluminum coating over an SiO$_{2}$ window. However the small particles of dust present over the inclined glass contaminated the spectra. Regular cleaning of the mask was required. It was replaced in 2019 by a true-hole mask fabricated in a Si wafer coated with protected aluminum. The wafers are as good as any optical mirror since they have an intrinsic flatness better than 2~$\mu$m over 150~mm and the rectangle, machined using etching, has errors of the order of a few microns. This new mask improved the  quality of the spectra.

\subsection{FOVSS}
\label{cdt:scans}
The FOVSS was designed to do 2D scans covering the total FOV up to $60^{\prime\prime}\times 60^{\prime\prime}$ \cite{Esteves18}. The scan size is configurable by the user. The FOVSS keeps the optical path length constant, regardless of the scanning position over the FOV. This condition is satisfied by means of three Scanning Mirrors (SM), which change their positions using three motors. Figure~\ref{cdt:path} includes the layout of the FOVSS configuration and its components. 
The entire assembly, with the exception of RS1, is mounted on a common translation stage (TS) which moves the system along the X-axis. The mask, plus the  SM1 and SM2 set are mechanically connected to a second TS, which moves this set along the Y-axis. Finally, the  SM2 and SM3 set is mechanically connected to a third TS which allows movement along the X-axis independently of the first TS. TS1 and TS2 allow the movement of the mask along the X and Y axes, respectively, to enable 2D scanning. TS3 uses the SM2 and SM3 set to compensate for the optical path length as a function of the system
movement while scanning. The total optical path length, determined by the sum of segments $b + c + d$ (shown in  Figure~\ref{cdt:path}), is always constant.

\begin{figure}[ht]
     \begin{center}
    \includegraphics[width=13cm]{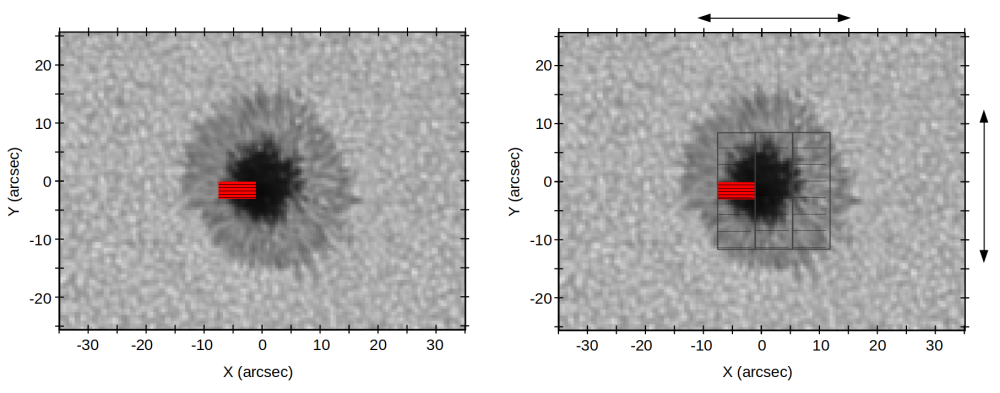}
      \caption{Example of the IFU mode in a single position (left frame). The image slicer is represented by a red rectangle (FOV of 6$^{\prime\prime}$~$\times$~3$^{\prime\prime}$). The right frame is an example of a 3~$\times$~7 scan (18$^{\prime\prime}$~$\times$~21$^{\prime\prime}$). The available scan movements are represented by the arrows, and a FOV up to $60^{\prime\prime}$~$\times$~$60^{\prime\prime}$ can be covered. The background image is from SDO/HMI continuum.}
         \label{cdt:figifu}
	 \end{center} 
   \end{figure}

Figure~\ref{cdt:figifu} shows a representation of the image slicer and some possible scan positions over the FOV, which is illustrated by a real-scale image from the Sun (SDO/HMI continuum). The image slicer is represented by a red rectangle, including the eight slicer mirrors. The left frame represents a single exposure over an area of 6$^{\prime\prime}$~$\times$~3$^{\prime\prime}$. The black rectangles in the right frame represent an example of a 3~$\times$~7 scan. The available scan movements are represented  by the arrows.

\begin{figure}
     \begin{center}
     \includegraphics[width=13cm]{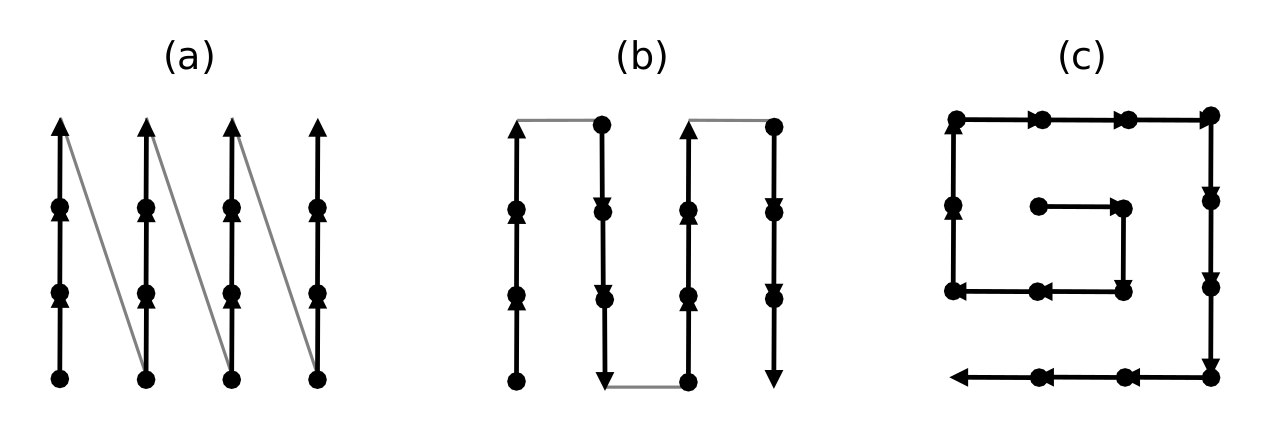}
        \caption{ Examples of the FOVSS scan patterns:  (a) Raster-vertical, (b) Snake-vertical and (c) Spiral-horizontal. The arrows represent the movements that define the scan positions.}
       \label{cdt:raster}
	 \end{center} 
   \end{figure}

The FOVSS allows scans to be made following different patterns, such as those shown in Figure~\ref{cdt:raster}. The patterns called ``-vertical'' are also available horizontally, to give transposed scanning patterns. 
 The initial scan position can be chosen between the geometrical center or one corner of the full scan pattern. Observers can choose the pattern that better fits the solar-feature dynamics they are studying.

\subsection{Performance of the IFU}

\begin{figure}[ht]
     \begin{center}
    \includegraphics[width=8cm]{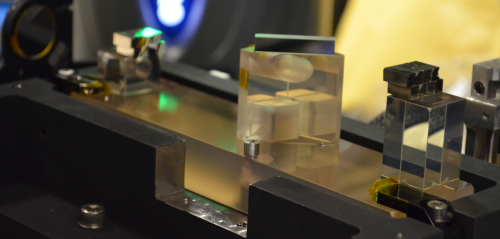}
    \includegraphics[width=8cm]{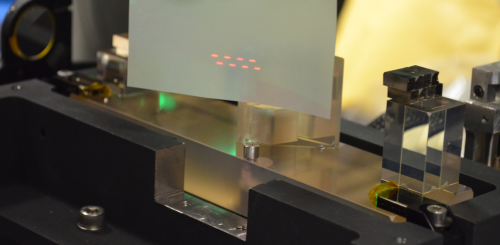}
      \caption{Pictures of the IFU in operation.  The IFU is the glass block (Zerodur). The visible elements in the left picture are, from left to right: the RS2 mirror, the array of collimator mirrors, the M1 folding mirror and the array of camera mirrors. The right picture shows the mini-slits projected onto a piece of paper. The polarimeter has been moved out of the light path to take the pictures.  }
         \label{cdt:pciifu}
	 \end{center} 
   \end{figure}

As described in Section~\ref{cdt:ifu}, the IFU has a set of collimator and camera mirrors to arrange the FOV into a slit. The mirror arrangement and numbering (illustrated in  Figure~\ref{cdt:numslice}) show that the central mirrors in the image slicer correspond to the edges of the collimator and mini-slit arrays. This layout produces some light contamination on the adjacent mirrors. Figure~\ref{cdt:pciifu} shows the IFU in operation and gives an idea of the small dimensions of its components.
 The size of each collimator and camera mirror is 4.2~mm~$\times$~4.5~mm.
  One possible source of  that contamination has been evaluated by studying the diffraction effects seen on the collimator mirrors due to the thin dimensions of the slicer mirror (100~$\mu$m).
 The tests were done at a wavelength of 15650~\AA, by illuminating only one slicer mirror at a time and measuring the light from all the output mini-slits. Table~\ref{cdt:tablextalk} shows the result of the tests. The measurements are represented as percentages, normalized to the self-illumination of each slice (diagonal values). The background level at the detector is subtracted. The maximum value of the slice-to-slice contamination  is 6.3\% for the neighbors of slice number 7, and the median value  is 3.5\%. The pupil mask reduces this value significantly. The same tests done without the mask give a median value of 4.8\%. Another effect, specific to the geometry of the mini-slits and collimator arrays (seen in Figure~\ref{cdt:figzem}), is ``intra-row'' light contamination, resulting from the layout of the collimator array in two rows. The values that are affected for each row are underlined (see  Table~\ref{cdt:tablextalk}). The maximum value of the intra-row contamination is 1.5\% from slice number 4 to number 8 and the median value  is 1.2\%. These measured values are compared with the modeled behavior (Regalado {\itshape et al.}, in prep.), which estimates that at least 80\% of the energy (from the first diffraction ring) is contained within a single slicer mirror. This value is lower than that obtained in the study by \citet{Calcines14}, because that work was done at a wavelength of 10000~\AA, which is a case with less diffraction. 
 An additional mask at the mini-slit plane could help to reduce some straylight in the direction perpendicular to the slits. The inclusion of this mask has been considered; the manufacturing and fixing, however, are difficult owing to the  discontinued form of mini-slits and the limited availability of space. It is planned for the future prototypes.

\begin{wstable}[ht]
    \caption{Light contamination measurements represented as percentages. Rows (I) denote the illuminated slicer mirror number  and columns (M) the measured mirror number. The values that are affected for each row are underlined. The numbering is shown in Figure~\ref{cdt:numslice}. }
     \label{cdt:tablextalk}
	 \begin{tabular}{c r r r r r r r r p{0.5\linewidth}} \toprule
        {\bf Mirrors}   &	 {\bf 4}  &  {\bf 3}  &  {\bf 2}  &  {\bf 1}  &  {\bf 5}  &  {\bf 6}  &  {\bf 7}  &  {\bf 8}  \\           
        {\bf (I $\setminus$ M)}   &	   &   &    &    &    &    &    &    \\ \colrule
	{\bf 4}  &  100  &  3.0  &  0.8  &  0.3  &  0.1  &  0.1  &  0.4  &  \underline{1.5}  \\
	{\bf 3}  &  3.5  &  100  &  3.3  &  1.0  &  0.5  &  0.5  &  \underline{1.1}  &  \underline{1.2}  \\
	{\bf 2}  &  0.5  &  3.4  &  100  &  3.3  &  0.9  &  \underline{1.5}  &  \underline{1.2}  &  0.3  \\
	{\bf 1}  &  0.1  &  0.8  &  3.4  &  100  &  \underline{4.2}  &  \underline{1.6}  &  0.5  &  0.3  \\
	{\bf 5}  &  0.0  &  0.2  &  0.8  &  \underline{3.5}  &  100  &  4.7  &  0.8  &  0.5  \\
	{\bf 6}  &  0.1  &  0.4  &  \underline{1.6}  &  \underline{1.5}  &  3.3  &  100  &  5.2  &  0.8  \\
	{\bf 7}  &  0.2  &  \underline{1.2}  &  \underline{1.2}  &  0.5  &  0.6  &  3.7  &  100  &  6.3  \\
	{\bf 8}  &  \underline{1.1}  &  \underline{1.1}  &  0.4  &  0.1  &  0.5  &  0.6  &  2.8  &  100  \\  \botrule
	 \end{tabular} 	 
   \end{wstable}

\section{Science observations}
\label{cdt:sciobsv}

Typical observations in IFU mode are similar to those carried out in slit mode: the observer has to choose the exposure time and the number of accumulations. The scan option  provides the possibility of covering a bigger FOV by controlling the number of scan steps. The only difference is that in IFU mode there are two dimensions for the movements, as explained previously.
 The scan can be automatically repeated in time to obtain  temporal series.
  In order to calibrate the observations, flat-field images have to be taken
     every
    1.5~h (maximum) and a series of polarimetric calibration images has to be recorded
    at least once per day.
     The latter is performed using the Polarimetric Calibration Unit \cite{Hofmann12}.

\begin{figure}
   \begin{center}
    \includegraphics[width=14cm]{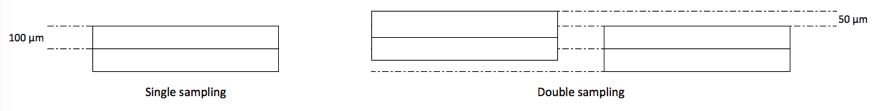}
    \caption {Representation of single sampling (left) and double sampling (right). The rectangles represent the slice mirrors and their apparent position with respect to the movements. Only two slice mirrors have been drawn for simplicity.}
     \label{cdt:doublesamp} 
     \end{center} 
\end{figure}

The spatial sampling can be done in single- or double-mode. Single sampling is equivalent to taking a single exposure at every position in the FOV. In double sampling mode, two exposures are taken at every position, one displaced half-slice-width with respect to the other. With it, the spatial resolution due to the sampling is doubled. The two spectral images are adequately interlaced by the reduction pipeline.
 Figure~\ref{cdt:doublesamp} illustrates how the double-sampling step size is 50~$\mu$m for the slice width of 100~$\mu$m ($0.375^{\prime\prime}$).\\

\begin{equation}
t_{\rm scan} \simeq [4 * nac *(t_{\rm int}+ t_{\rm readout})+ t_{\rm mov}]* sampling * nv * nh
\label{cdt:eqtscan}
\end{equation}

The scan time can be estimated using Eq.~\ref{cdt:eqtscan}, where: \\

\noindent $nac$ = number of accumulations\\
$t_{\rm int}$ = integration time\\
$t_{\rm readout}$ = 30 ms\\
$t_{\rm mov}$ = 1.3 s approx.\\
$sampling$ = 1 for single, 2 for double sampling (see the text)\\
$nv$ =  number of vertical steps (running perpendicular to the slices)\\
$nh$ =  number of horizontal steps (running parallel to the slices)\\

   \begin{figure}[ht]
     \begin{center}
    \includegraphics[width=8cm]{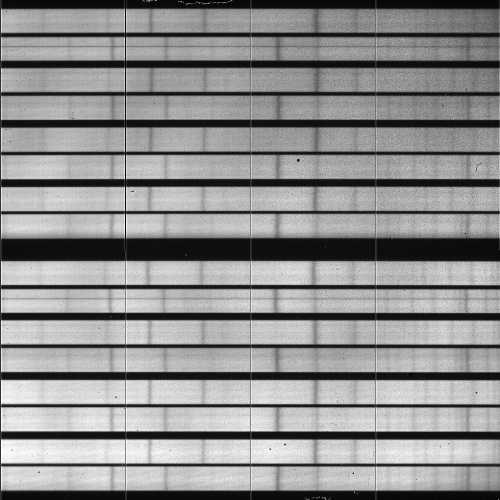}
      \caption{Example of the raw image of IFU spectra in spectropolarimetric mode centered at 15650~\AA. The wavelength range is 40~\AA  $\:$ and Fe~{\sc i} absorption lines are visible. The most intense one is Fe~{\sc i}~15662.0~\AA. Two series of eight spectra are shown for two orthogonal states of polarization, one at the top of the raw image and the other at the bottom. The apparent shift in wavelength between two consecutive spectra is an optical shift due to the arrangement of the mini-slits in two rows. }
         \label{cdt:figdet}
	 \end{center}
   \end{figure}

\subsection{Reduction pipeline}

 The reduction pipeline \cite{Collados03} was updated to recognize the IFU mode and to calibrate the data in any observing  mode. It is based on the  procedures described in \citet{Collados99}. 
The current version, GRIS\_V8 is used for the data reduction process with both slit and IFU modes.
 The pipeline does the dark current and flat corrections, as well as the wavelength and polarimetric  calibrations. The pipeline also reorganizes the image of the spectra, from what is seen in the detector  (see Figure~\ref{cdt:figdet}) into a 3D data cube  containing two  spatial directions plus the spectral direction. The resulting file contains  two additional dimensions, one for the Stokes parameters ($I$, $Q$, $U$, and $V$) and one for time, in the case of temporal series. The data has a spatial sampling of $0.135^{\prime\prime}\times~0.1875^{\prime\prime}$, where the first number is the pixel scale and the second is, in double sampling mode, the slice width divided by 2. In single sampling mode it is $0.135^{\prime\prime}\times~0.375^{\prime\prime}$.

 Figure~\ref{cdt:figdet} shows an example of the raw image of IFU spectra  seen at the detector. It was taken in spectropolarimetric  mode, so there are two groups of eight spectra, corresponding to two orthogonal states of polarization. 
 The central wavelength is 15650~\AA $\:$ and the  range is 40~\AA. There is an apparent shift in wavelength between two  consecutive spectra. This effect is actually an optical shift due to the arrangement of the mini-slits in two rows (as seen in the insert of Figure~\ref{cdt:figzem}). Some  bad pixels and the readout structure (vertical lines) of the detector are also  present in this raw image. The  pipeline removes all these artifacts and corrects the optical shift. Five spectral absorption lines are visible,  Fe~{\sc i}~15662.0~\AA $\:$  being the most intense one.  The wavelengths of the other Fe~{\sc i} absorption lines are: 15645.0~\AA, 15648.5~\AA, 15652.9~\AA $\:$ and 15665.2~\AA.

\subsection{Examples of observations}

Some examples of the data obtained with the IFU are described below in order to show the instrument capabilities, versatility and performance.

\subsubsection{1~$\times$~2 scan of solar granulation}   

\begin{figure}[ht]
    \begin{center}
       \includegraphics[width=17.9cm]{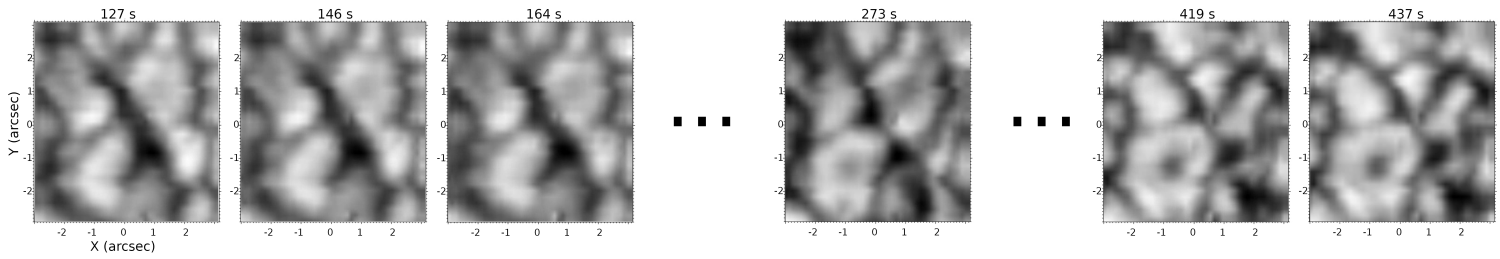}
  \caption{Example of a series of continuum intensity  images near 15650~\AA $\:$ of the solar granulation, observed with a 1~$\times$~2 scan ($6^{\prime\prime}\times~6^{\prime\prime}$). The cadence is 18~s and the elapsed time relative to the start of this temporal series is printed at the top of each image. }
    \label{cdt:figeg}
     \end{center}
\end{figure}

A small region at the center of the solar disk was observed after the commissioning on September 2018.
Figure~\ref{cdt:figeg} shows an example of the time-series observations with a 1~$\times$~2 scan ($6^{\prime\prime}\times~6^{\prime\prime}$). This series shows the reconstructed images of the continuum intensity near 15650~\AA , observed with a cadence of 18~s. The total duration of the series is about 90 minutes. The evolution of the granulation is clearly seen over the selected time range for the figure. 
  The number at the top of each image is the elapsed time relative to the start of this temporal series.  The formation of one exploding granule is distinguishable at the lower-left corner of the image at 273~s, and its evolution can be followed over the subsequent images. The analysis of the evolving magnetic fields is in progress  (Dominguez-Tagle {\itshape et al.}, in prep.).

\begin{figure}
    \begin{center}
       \includegraphics[width=15cm]{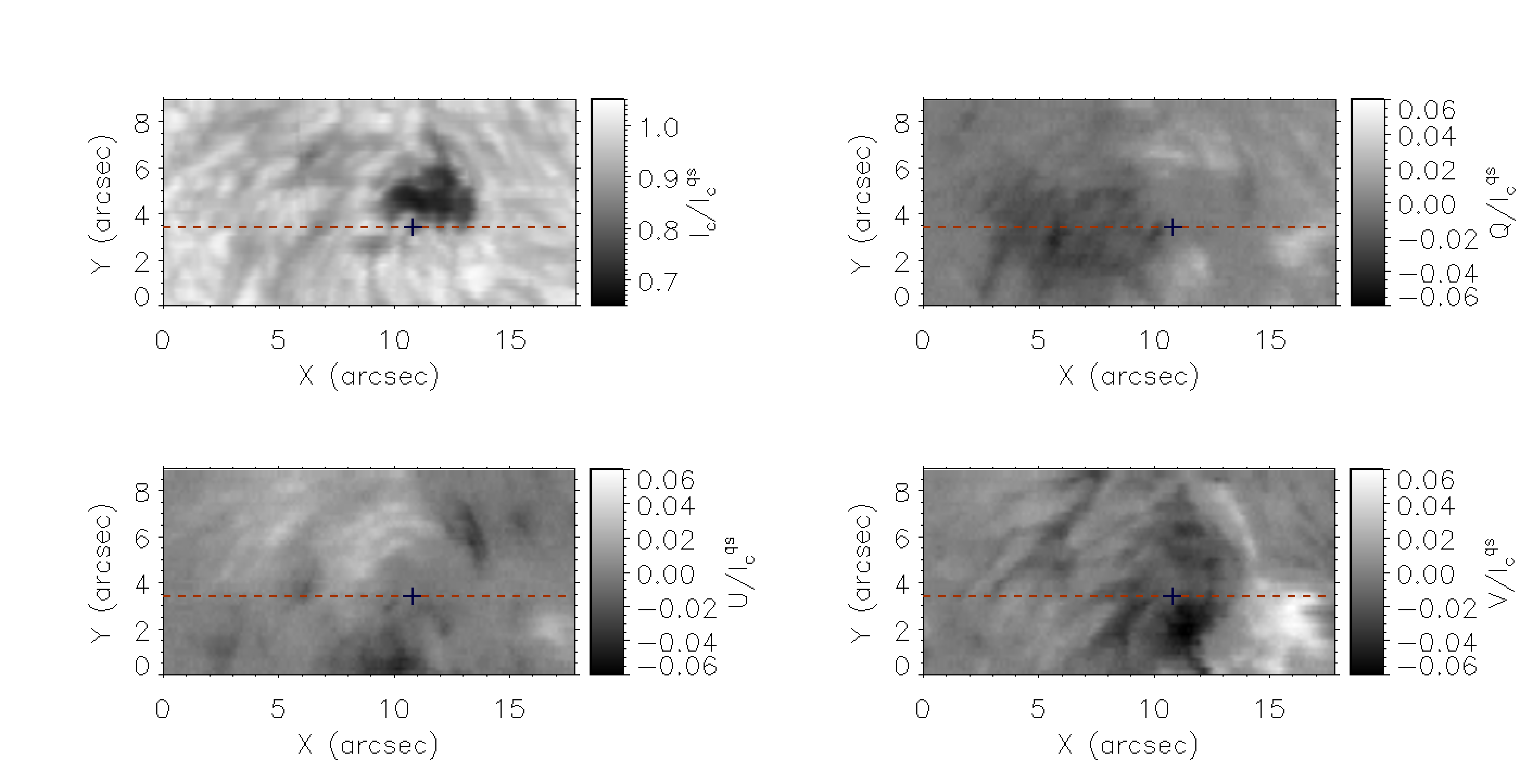}
    \caption {Monochromatic images of parts of active region 12665 observed with a 3~$\times$~3 scan ($18^{\prime\prime}\times~9^{\prime\prime}$). The images are, from left to right and top to bottom, the continuum intensity and the Q, U, V monochromatic images at a wavelength located 0.5~\AA $\:$ to the blue from the center of the  Fe~{\sc i}~15648.52~\AA $\:$ spectral line. }
        \label{cdt:monoimg}
    \end{center}
\end{figure}

\subsubsection{3 $\times$ 3 scan of solar active region}

Active region 12665 (located at solar coordinates S06E24) was observed on July 2017, with a 
3~$\times$~3 scan ($18^{\prime\prime}\times~9^{\prime\prime}$). It was observed over 24 minutes in a temporal series with  a cadence of 40~s.  This FOV covers a region with a small umbra with a spiral-shaped penumbra.
Figure~\ref{cdt:monoimg} shows the continuum intensity map and the Q, U, V monochromatic images extracted from the polarized spectrum at 0.5~\AA $\:$ to the blue from the center of the  Fe~{\sc i}~15648.52~\AA $\:$  spectral line. This spectral line is specially suited  for magnetic studies, because of its large magnetic sensitivity (Land\'e factor $g_{\rm eff}=2$). The data sets have been obtained with three modulation cycles. In each of these, a set of four images is recorded, from which the four Stokes parameters I, Q, U, and V can be retrieved. All images corresponding to the same modulation step are summed up in real time. 
The signal to noise ratio achieved a value of 700 in the Q, U, and V continua ($I_c/\sigma_{Q,U,V}$, where $I_c$ 
is the continuum intensity and  $\sigma_{Q,U,V}$ is the Q, U, V continuum noise level -one standard deviation-).

\begin{figure}
    \begin{center}
       \includegraphics[width=15cm]{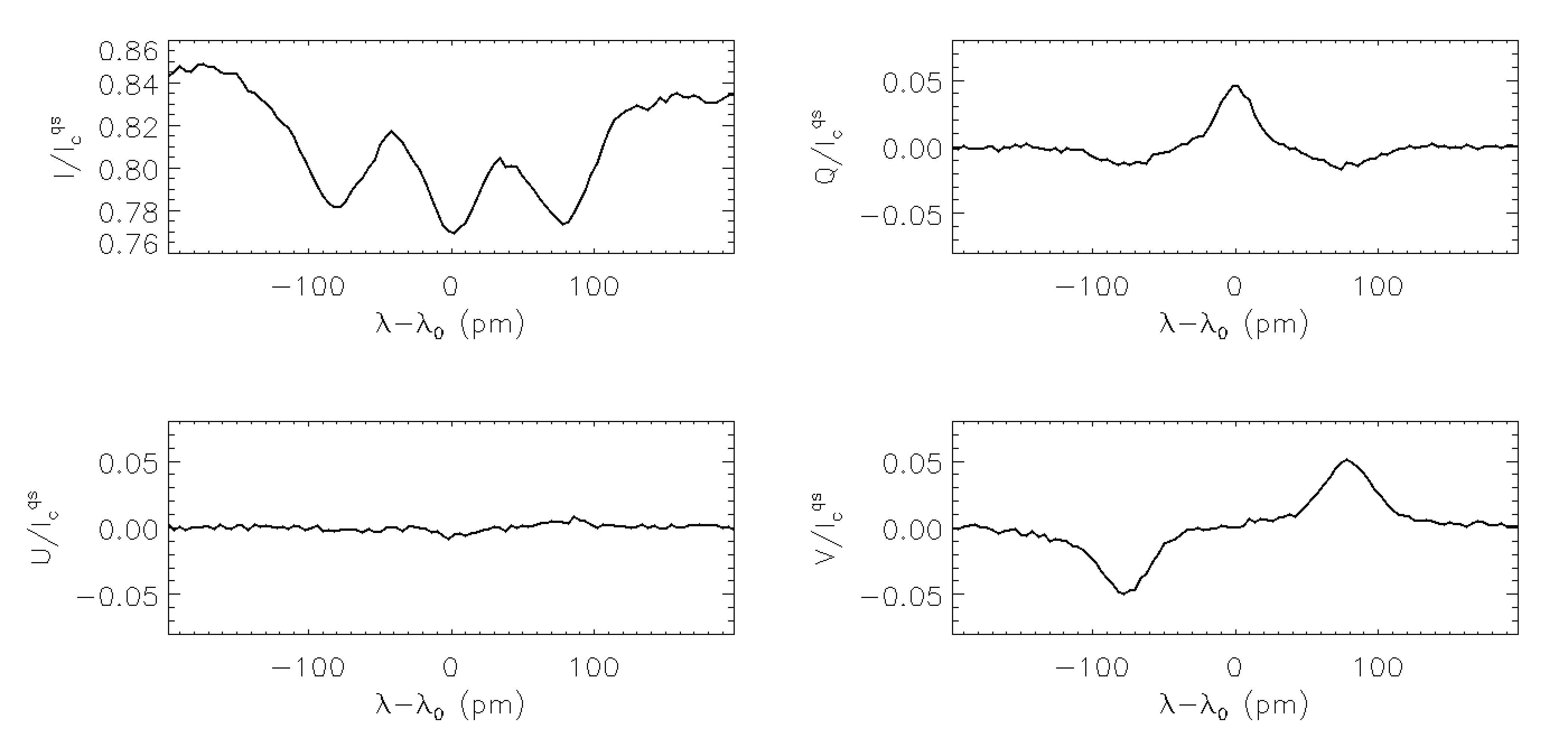}
    \caption {Example of polarized spectral profiles in a small wavelength range  around the center of the Fe~{\sc i}~15648.52~\AA $\:$ spectral line ($\lambda_0$). The images correspond to the points indicated with a blue cross in Fig.~\ref{cdt:monoimg}. The Zeeman splitting of the intensity profiles is apparent, as well as the amplitude and separation of the Zeeman components in the polarized profiles.}
    \label{cdt:specprof}
     \end{center}
     \end{figure}

\begin{figure}
    \begin{center}
       \includegraphics[width=15cm]{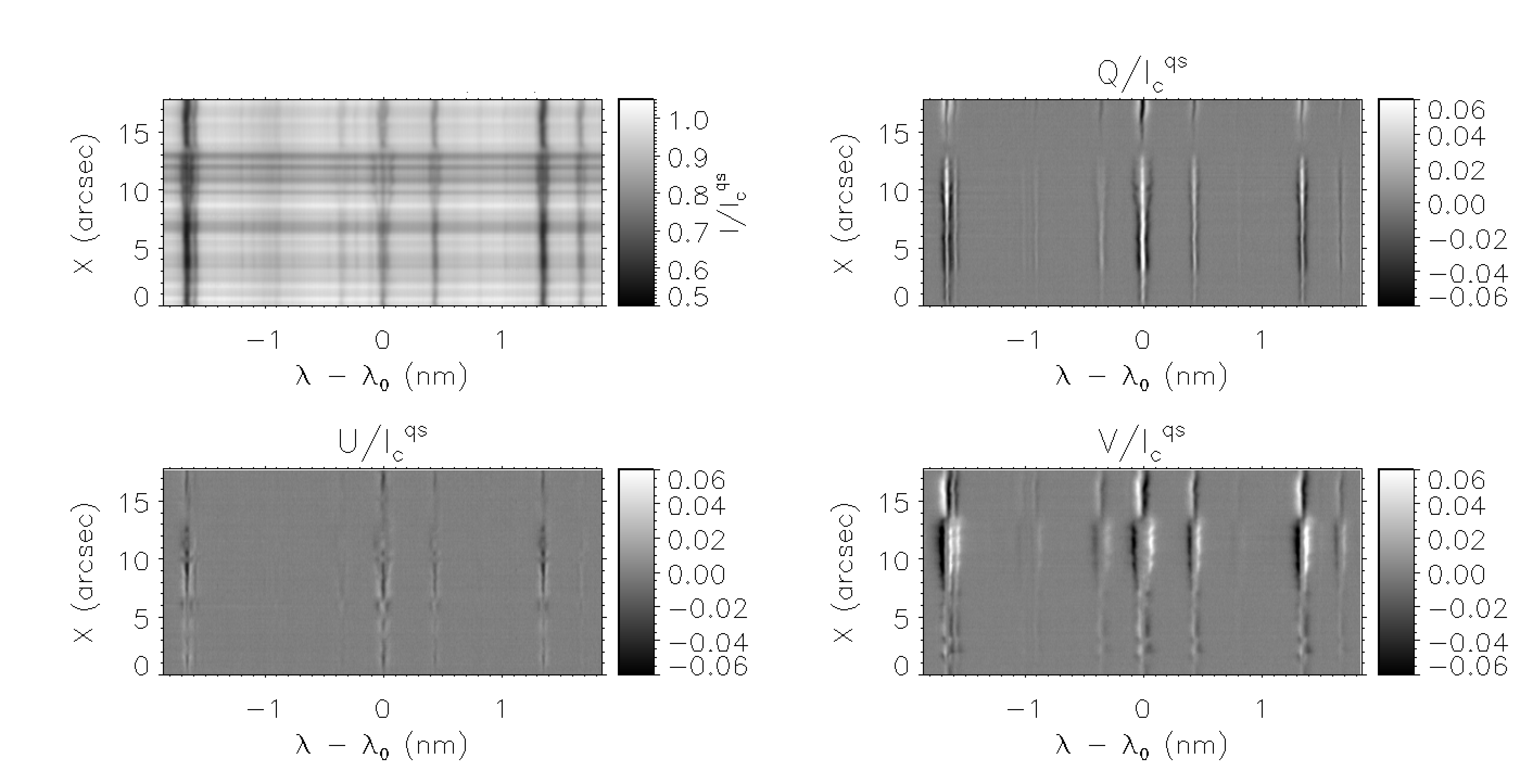}
    \caption {Equivalent horizontal long-slit spectral images, corresponding to the red dashed lines displayed
    in Fig.~\ref{cdt:monoimg}. The four images displays, from left to right and top to bottom, 
    the I, Q, U, and V profiles, normalized to the continuum intensity of the quiet Sun at the
    solar disk center. The X-axis is centered on the wavelength of the Fe~{\sc i}~15648.52~\AA $\:$ spectral line ($\lambda_0$), and the Y-axis represents the horizontal length of each map.}
        \label{cdt:specimg}
    \end{center}    
\end{figure}

Figure~\ref{cdt:specprof} shows the I, Q, U, V profiles at the point indicated by the blue cross in Figure~\ref{cdt:monoimg}. Only a spectral range of 4~\AA $\:$ around the Fe~{\sc i}~15648.52~\AA $\:$ spectral line is shown, whereas the recorded spectral range  is 40~\AA. The Zeeman splitting in these selected points corresponds to about  2300~G. The amplitudes of the linearly polarized profiles indicate a non-negligible inclination of the magnetic field.
The red dashed lines in Figure~\ref{cdt:monoimg} are used to display equivalent horizontal long-slit I, Q, U and V spectral images (see Figure \ref{cdt:specimg}). The spatial variations of the field strength, polarity and orientation are clearly detected, as indicated by the separation and sign of the Zeeman components in the Q, U and V images.

\subsubsection{Published results based on IFU observations}

\begin{itemlist}
\item Quiet-Sun magnetic flux cancellations  were studied by \citet{Anjali20}, from  observations of quiet-Sun granulation performed on November 2018. The 40-minute  time-series observations have a cadence of 26.4~s, for a 1~$\times$~2 scan ($6^{\prime\prime}\times~6^{\prime\prime}$). The spectral range was centered on the Si~{\sc i}~10827.108~\AA $\:$ spectral line, which is magnetically sensitive with a Land\'e factor $g_{\rm eff}=1.5$. \\

\item A plage region in active region 12723 was observed on October 2018, and the analysis of the magnetic field structures, published by \citet{tetsu21} is partially based on observations with GRIS/IFU. Three temporal series of 1~$\times$~2 scan ($6^{\prime\prime}\times~6^{\prime\prime}$) with a cadence of 26~s were performed by the IFU. The  observations were centered on a spectral region containing the He~{\sc i} triplet at 10830~\AA $\:$  and Si~{\sc i}~10827.108~\AA $\:$ spectral line.\\

\item Magnetic fields in photospheric small-scale network regions were studied by \citet{campbell21}, from observations  performed on May 2019.  A 3~$\times$~3 scan ($18^{\prime\prime}\times~9^{\prime\prime}$) of quiet Sun inter-network regions, very close to the disk center, was used for the observations. Their analysis is based on five spectral absorption lines, including Fe~{\sc i}~15648.52~\AA $\:$  and Fe~{\sc i}~15652.87~\AA .\\

\item On September 2019, a 3~$\times$~3 scan ($18^{\prime\prime}\times~9^{\prime\prime}$) over the lead pore of active region 12748, was conducted by \citet{nelson21} to study the line-of-sight magnetic field strength in pores. Their research is mostly based on a GRIS/IFU time-series observations of Fe~{\sc i}~15648.52~\AA $\:$  and Fe~{\sc i}~15652.87~\AA $\:$  spectral lines, with a cadence of 67~s.

\end{itemlist}

\section{Conclusions}
\label{cdt:concl}

An IFU was designed and successfully commissioned on 
 the GRIS spectrograph installed at the GREGOR solar telescope of the Observatorio del Teide (Tenerife).
 The IFU, based on image slicers, opens up new possibilities for NIR observations of fast-evolving features in the Sun. It has  very good optical quality with low straylight, and  gives unique performance, allowing faster small-area scans than a traditional slit spectropolarimeter.

  Based on the success of the first campaigns, the IFU has been offered  to all observers who can be granted telescope observing time.
  So far, more than two years of operations have been completed and more than 15 teams of scientists have observed with this instrument. The response of  observers has been very positive in terms of both feedback comments and the high demand for the instrument. The first scientific results of observations with the IFU are beginning to be published.

\subsection* {Acknowledgments}
This work was carried out with the funding of the Projects SOLARNET (FP7; funded by the European Commission's 7th Framework Program under grant agreement no. 312495), GREST (funded by the European Commission's H2020 Program under grant agreement no. 653982) and SOLARNET (H2020; funded by the European Commission's H2020 Program under grant agreement no. 824135). SDO data are provided by the Joint Science Operations Center-Science Data Processing. 

The 1.5 m GREGOR solar telescope was built by a German consortium under the leadership of the Leibniz-Institut f\"{u}r Sonnenphysik in Freiburg with the Leibniz-Institut f\"{u}r Astrophysik Potsdam, the Institut f\"{u}r Astrophysik G\"{o}ttingen, and the Max-Planck-Institut f\"{u}r Sonnensystemforschung in G\"{o}ttingen as partners, and with contributions by the Instituto de Astrof\'{i}sica de Canarias and the Astronomical Institute of the Academy of Sciences of the Czech Republic. 

\vspace{0.5cm}

Preprint of an article submitted for consideration in Journal of Astronomical Instrumentation © 2022 [copyright World Scientific Publishing Company] https://doi.org/10.1142/S2251171722500143

\bibliography{dom22}   
\bibliographystyle{ws-jai}

\end{document}